\documentclass[sigconf,nonacm]{acmart}

\AtBeginDocument{%
  }

\newcommand{\newterm}[1]{\textbf{\textit{#1}}}

\usepackage{natbib}
\usepackage{arydshln}
\usepackage{subfigure}
\usepackage{balance}
\usepackage{enumitem}
\usepackage{pifont}
\usepackage[dvipsnames]{xcolor}

\usepackage{booktabs, multirow}

\usepackage{framed}

\usepackage{algorithm}
\usepackage{algorithmic}

\begin{document}

\title{Credit Network Modeling and Analysis via Large Language Models}

\author{Enbo Sun}
\email{sgesun@liverpool.ac.uk}
\affiliation{
  \institution{University of Liverpool}
   \city{Liverpool}
  \country{UK}
}

\author{Yongzhao Wang}
\email{yongzhao.w@unf.edu}
\affiliation{
  \institution{University of North Florida}
   \city{Jacksonville}
  \country{USA}
}

\author{Hao Zhou}
\email{haozhou.ai@gmail.com}
\affiliation{%
  \institution{Nanchang University}
   \city{Nanchang}
  \country{China}
}

\begin{abstract}

We investigate the application of large language models (LLMs) to construct credit networks from firms' textual financial statements and to analyze the resulting network structures.
We start with using LLMs to translate each firm's financial statement into a credit network that pertains solely to that firm.
These networks are then aggregated to form a comprehensive credit network representing the whole financial system. 
During this process, the inconsistencies in financial statements are automatically detected and human intervention is involved.
We demonstrate that this translation process is effective across financial statements corresponding to credit networks with diverse topological structures.
We further investigate the reasoning capabilities of LLMs in analyzing credit networks and determining optimal strategies for executing financial operations to maximize network performance measured by the total assets of firms, which is an inherently combinatorial optimization challenge.
To demonstrate this capability, we focus on two financial operations: portfolio compression and debt removal, applying them to both synthetic and real-world datasets.
Our findings show that LLMs can generate coherent reasoning and recommend effective executions of these operations to enhance overall network performance.

\end{abstract}

\maketitle

\section{Introduction}

Financial systems rely heavily on understanding complex relationships between entities, such as borrowers, lenders, and intermediaries. 
Traditionally, these relationships are modeled using structured data, such as balance sheets and transaction records. 
However, a significant portion of financial knowledge exists in unstructured natural language including contracts, reports, and regulatory filings, making automated analysis challenging. 
Recent advances in \newterm{large language models} (LLMs) present an opportunity to bridge this gap by extracting and formalizing financial relationships from text into structured representations.  Once structured, LLMs' reasoning capabilities may further support diverse analyses of the financial system and contribute to its overall stability and robustness.

In this study, we develop LLM-based tools to construct network representations of financial systems via the financial statements of their \newterm{firms} and to analyze the resulting network structures.
We first introduce a framework that translates the financial statements into \newterm{credit networks}, which are graph-based representations of financial systems, with nodes representing firms and edges capturing debt relationships. 
Credit networks have been widely used to study various aspects of financial systems, such as assessing their stability~\cite{maeno2012transmission} and forecasting the impact of financial operations on the broader system~\cite{schuldenzucker2021portfolio,veraart2022does,mayo2021strategic}.
By automating the construction of credit networks from large volumes of unstructured financial statements, our framework can substantially reduce manual effort and enable more efficient analysis.

While we permit financial statements to be unstructured, constructing a credit network requires that they include specific information: the firm's debt relationships, debt amounts, and assets such as bank balances and treasury bills.
In essence, a financial statement can be interpreted as a natural language description of a \newterm{balance sheet}, adhering to the conventions of \emph{Generally Accepted Accounting Principles} (GAAP) and \emph{International Financial Reporting Standards} (IFRS).
Given such input, we prompt LLMs to generate a credit network for each firm that pertains solely to that firm. 
These individual credit networks are then combined into aggregated credit networks that represent the structure of the overall financial system. 
During this process, the discrepancies in financial statements are automatically detected and human intervention is required. 
We evaluate our framework across various translation tasks that differ in the content of financial statements and the resulting network topologies. 
Across all tasks, the framework consistently produces correct translations, demonstrating its robustness and reliability.

The second objective of this study is to explore how LLMs can be used to analyze credit networks and enhance their performance, particularly by increasing the total assets held by firms. 
Prior research typically tackles this goal through \newterm{financial operations} such as debt transfers~\cite{kanellopoulos2023debt}, \newterm{portfolio compression}~\cite{schuldenzucker2021portfolio,veraart2022does,mayo2021strategic}, and prepayments~\cite{hao2024prepay}, which aim to restructure debt relationships to improve overall network efficiency. 
However, identifying the optimal execution of these operations (e.g., selecting which \newterm{debt cycles} to compress) presents a combinatorial optimization problem. 
A poor execution can destabilize the system and trigger cascades of bankruptcies.
To tackle this challenge, we prompt the LLM to determine the most effective execution strategies. 
Each prompt includes a target network, a description of the financial operation, a clearing algorithm, a performance measure, and a task description.
We demonstrate this approach using two financial operations: portfolio compression and \newterm{debt removal}~\cite{papp2020network,kanellopoulos2022forgiving,tong2024reducing}. 
For both operations, our framework enables the LLM to reason about execution strategies that significantly improve total system assets compared to random or no execution as well as heuristic execution strategies.

The contributions of this work include:

\begin{enumerate}
    \item A framework that converts financial statements into credit networks using LLMs;
    \item A framework that leverages the reasoning capabilities of LLMs to guide the execution of financial operations within credit networks;
    \item A demonstration of the effectiveness of our frameworks in various financial systems.
\end{enumerate}

\section{Related Work} \label{sec: related work}

Recent advancements in LLMs have generated increasing interest in their applications to finance, leading to a growing body of research focused on adapting and evaluating these models in financial contexts.

One of the earliest domain-specific models, \newterm{FinBERT}, was introduced by \citet{huang2023finbert}. Pre-trained on financial texts such as earnings calls, 10-K filings, and analyst reports, FinBERT demonstrated superior performance over general-purpose LLMs and traditional machine learning approaches in tasks including sentiment analysis, ESG classification, and financial tone detection, particularly in low-resource settings.
Building on this trend, \citet{wu2023bloomberggpt} presented \newterm{BloombergGPT}, a large-scale model trained on a vast mix of proprietary financial and general data. BloombergGPT supports a wide range of financial natural language processing tasks, such as named entity recognition, sentiment analysis, question answering, document summarization, and news classification. Integrated within Bloomberg's internal systems, it provides real-time market insights, data-to-text generation, and chatbot support, showcasing the potential of domain-adapted LLMs in production environments.
Similarly, \citet{xingye2025measurement} applied LLMs to analyze corporate annual reports, uncovering patterns of digital transformation in financial systems.
Other researchers have explored the fine-tuning of general LLMs on financial data. For example, \citet{xie2023pixiu} introduced \newterm{FinMa}, a fine-tuned variant of LLaMA~\cite{touvron2023llamaopenefficientfoundation}, tailored for finance. Their empirical evaluation highlighted the strengths and weaknesses of FinMa across various financial tasks.

Recognizing the need for systematic evaluation, \citet{xie2024finben} developed \newterm{FinBen}, a comprehensive financial benchmark comprising 36 datasets spanning 24 tasks across seven critical dimensions: information extraction, textual analysis, question answering, text generation, risk management, forecasting, and decision-making.
In the context of financial reporting, \citet{yang2024evaluating} evaluated state-of-the-art LLMs on summarizing financial statements, measuring their performance in terms of accuracy, informativeness, and coherence.

Several studies have also focused on the challenges LLMs face in financial domains. \citet{han2024xbrl} highlighted the limitations of current models in analyzing financial statements encoded in the eXtensible Business Reporting Language (XBRL). To address these challenges, they introduced \newterm{XBRL-agent}, which leverages retrieval-augmented generation and task-specific tools for improved performance on structured financial documents.

To support future research, survey efforts by \citet{lee2025large} and \citet{nie2024survey} have provided overviews of the LLM-for-finance landscape. The former focused on model training paradigms, while the latter categorized applications across linguistic processing, sentiment analysis, time series modeling, financial reasoning, and agent-based simulations.

Finally, \citet{liu2023fingpt} proposed \newterm{FinGPT}, a framework for automating the collection and curation of real-time financial data from the Internet. This system facilitates the fine-tuning of LLMs, supporting up-to-date and context-aware financial modeling.

\section{Preliminaries}

\subsection{Credit Networks}

\begin{figure}[!h]
    \centering
    \includegraphics[width=\linewidth]{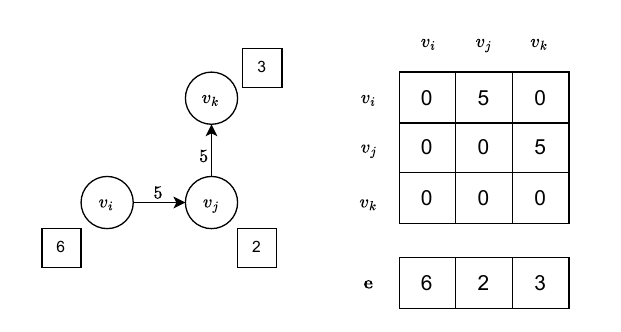}
    \caption{An example of a credit network. The credit network consists of three firms $v_i$, $v_j$, and $v_k$ with external assets of 6, 2, and 3, respectively. Firm $v_i$ owes 5 to $v_j$, and firm $v_j$ owes 5 to $v_k$. Liability matrix $\mathbf{L}$ and external assets $\mathbf{e}$ are shown on the right-hand side.}
    \label{fig: credit network example}
\end{figure}

A credit network is a graph, denoted as $G=(V,E)$, where nodes represent firms, and edges symbolize debt contracts. 
We denote the number of firms by $n = |V|$, and represent the set of firms as $[n] := \{1, \dotsc, n\}$.
Each firm $v_i \in V$ with $i\in[n]$ initially holds non-negative \newterm{external assets} $e_i \in \mathbb{R}_{\geq 0}$, representing liquid assets received from entities external to the financial system. 
Moreover, firms are connected by directed edges that represent \newterm{liabilities}. 
The directed edge $(v_i, v_j) \in E$ signifies a financial relationship between firms $v_{i}$ and $v_{j}$, with $l_{ij} \geq 0$ denoting the liability that firm $v_{i}$ (the borrower) owes to firm $v_{j}$ (the lender). 
Both $l_{ij} > 0$ and $l_{ji} > 0$ might coexist. 
If no directed edge connects firms $v_{i}$ and $v_{j}$, then $v_{i}$ has no liability to $v_{j}$, and $l_{ij} = 0$. 
The graph $G$ is irreflexive, meaning no firm can be liable to itself, resulting in a directed and positively weighted graph without self-loops. 
Therefore, a credit network can be represented by a liability matrix $\mathbf{L} \in \mathbb{R}^{n \times n}$ along with a vector of external assets $\mathbf{e} \in \mathbb{R}^{n}$.
An example of a credit network and its matrix representation are shown in Figure~\ref{fig: credit network example}.

Let $p_{ij}$ denote the actual payment made by firm $v_{i}$ to firm $v_{j}$, where $p_{ii} = 0$ and $p_{ij}$ may not be equal to the liability $l_{ij}$ since $l_{ij}$ may not be affordable for firm $i$. 
We denote by $\mathbf{P} = (p_{ij})$ for $i,j \in [n]$ the \newterm{payment matrix}.
The total outgoing payments of $v_i$ are represented by $p_i = \sum_{j \in [n]}{p_{ij}}$. 
Given a payment matrix $ \mathbf{P} $, the \newterm{total assets} of firm $ i $, $ a_i(\mathbf{P}) $, comprise its external assets and incoming payments, that is, $ a_i(\mathbf{P}) = e_i + \sum_{j \in [n]} p_{ji} $.
The \newterm{equity} is defined as $ E_i(\mathbf{P})=\max\{0,\ a_i(\mathbf{P})-L_i\}$ where $L_i=\sum_{j \in [n]} {l_{ij}}$ is the total liability of $v_i$.
Firms capable of fully meeting their obligations are considered \newterm{solvent}, that is, $a_i(\mathbf{P}) \ge L_i $; Otherwise, they are labeled \newterm{in default} or \newterm{insolvent}.
In the case of insolvency, a firm may need to liquidate its assets or make payments beyond the financial system, such as salary disbursements. 
Consequently, a default firm can only utilize an $\alpha\in [0,1]$ fraction of its external assets and incoming payments. 
When $\alpha = 1$, it means an absence of default costs.
We omit the dependence on $\mathbf{P}$ when the payments are clear from the context.

Payments $\mathbf{P}$ are determined using \newterm{clearing algorithms}.
In this study, we consider maximal clearing payments, which maximize each individual payment component-wise.
These payments are guaranteed to exist, are unique, and can be computed in polynomial time~\cite{rogers2013failure}.
The clearing algorithm that produces these payments operates and according to the following rules.
Specifically, a solvent firm must fully pay all its obligations, while a firm in default can only pay partially but should repay as much of its debt as possible, based on its total assets affected by the default costs. 
A partial payment to a lender should be proportional to its liability to the same lender. 
Mathematically, the payment from $v_i$ to $v_j$ with $i,j \in [n]$ satisfies $p_{ij} = l_{ij}$ when $v_i$ is solvent, and $p_{ij} =\alpha \cdot \left( e_i+ \sum_{j=1}^n x_{ji}\right) \cdot \frac{l_{ij}}{L_i}$ when $v_i$ is in default. 
We set $\alpha=0.5$ in the clearing algorithm for our setup.

\subsection{Financial Operations}

A financial operation refers to any activity or transaction within a financial system, undertaken by individuals, firms, financial institutions, or governments to achieve specific economic or strategic objectives.
Financial operations can be naturally represented within credit networks.
Here we introduce two financial operations that we use to demonstrate our approach.

\paragraph{\textbf{Portfolio Compression:}}

Portfolio compression is an operation that reduces the debt cycles in a financial system.
By shrinking the interdependencies of firms in the system, portfolio compression can limit the channels through which financial distress can propagate in the event of a firm’s insolvency and stabilize the system.

\begin{figure}[!htbp]
    \centering
    \includegraphics[width=0.9\linewidth]{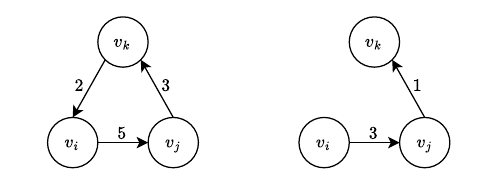}
    \caption{An example for portfolio compression. \textbf{Left:} An initial credit network. \textbf{Right:} An updated credit network after portfolio compression with $\mu^c=2$.}
    \label{fig: portfolio compression}
\end{figure}

In a credit network, portfolio compression of a directed debt cycle $c$ reduces each liability along the cycle by $\mu^c$, yielding updated liabilities $l^c_{ij} = l_{ij} - \mu^c$ for all $(v_i, v_j) \in c$, where $\mu^c = \min_{(v_i, v_j) \in c} l_{ij}$ denotes the minimum liability on the cycle.
An example of portfolio compression is shown in Figure~\ref{fig: portfolio compression}.

When compressing multiple debt cycles simultaneously, the order of compression is crucial, as one compression can eliminate edges or reduce liabilities in other cycles.
To address this, we use a tie-breaking rule: debt cycles involving the most firms are given highest priority for compression.
If two cycles involve the same number of firms, the order of compression between them is chosen at random.

\paragraph{\textbf{Debt Removal:}}

Debt removal in a financial system involves reducing or eliminating outstanding debt obligations through voluntary repayment, negotiated solutions such as restructuring or forgiveness. 
This process can help stabilize the financial system by restoring the solvency of bankrupt firms. 
By restoring solvency, the negative effects of defaults are mitigated, leading to an increased payment flow across the system, which, in turn, can benefit firms that forgive the debt.

\begin{figure}[!htbp]
    \centering
    \includegraphics[width=0.9\linewidth]{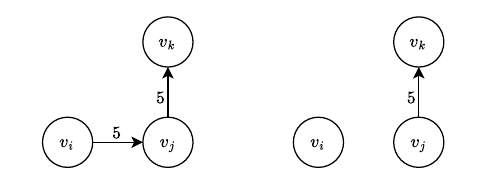}
    \caption{An example for debt removal. \textbf{Left:} An initial credit network. \textbf{Right:} An updated credit network after removing debt $l_{ij}$.}
    \label{fig: debt removal}
\end{figure}

In a credit network, debt removal is a financial operation that eliminates an existing edge $(v_i, v_j)$ from the network, thereby removing the liability $l_{ij}$ owed by borrower $v_i$ to lender $v_j$.
An illustration of this process is provided in Figure~\ref{fig: debt removal}.

Financial operations often admit combinatorially many execution strategies. For example, in portfolio compression, one can choose among any subset of debt cycles to compress. 
However, not every execution is beneficial to the financial system. 
In some cases, such as compressing certain debt cycles, the result can be destabilizing (e.g., triggering cascades of bankruptcies) as shown by~\citet{schuldenzucker2020default}. 
Consequently, it is essential to identify execution strategies that optimize specific systemic criteria, such as the sum of total assets (i.e., $\sum_{i \in [n]} a_i$). 
Yet, finding such an optimal execution is a challenging problem of combinatorial optimization.

\subsection{In-Context Learning}

Our methods are based on \newterm{in-context learning}.
In-context learning refers to the ability of an LLM to learn new information or skills by
observing examples or instructions provided in its input, without any additional training or
fine-tuning~\cite{dong2022survey}.

Consider an LLM represented as a function $\mathcal{M}$, trained to predict the next token probabilistically. Given an input prompt $z = (z_1, z_2, \dotsc, z_n)$, the model assigns probabilities to possible next tokens $y$ as:
\[
\mathcal{M}(y|z) = P(y|z_1, z_2, \dotsc, z_n).
\]

In in-context learning, the input prompt includes a sequence of example pairs $(Q_i, A_i)$, which may be question-answer pairs or task demonstrations. These examples serve as context for the model to generate an answer $A_{n+1}$ for a new query $Q_{n+1}$. Formally, the model's prediction is given by:
\[
\mathcal{M}(A \,|\, (Q_1, A_1), (Q_2, A_2), \dotsc, (Q_n, A_n), Q_{n+1}).
\]

Crucially, in-context learning does not involve updating the model's weights. Instead, the LLM leverages the provided examples within its context window to adjust its conditional probability distribution for the next token.

\begin{figure*}[!t]
  \centering
  \includegraphics[width=0.85\textwidth]{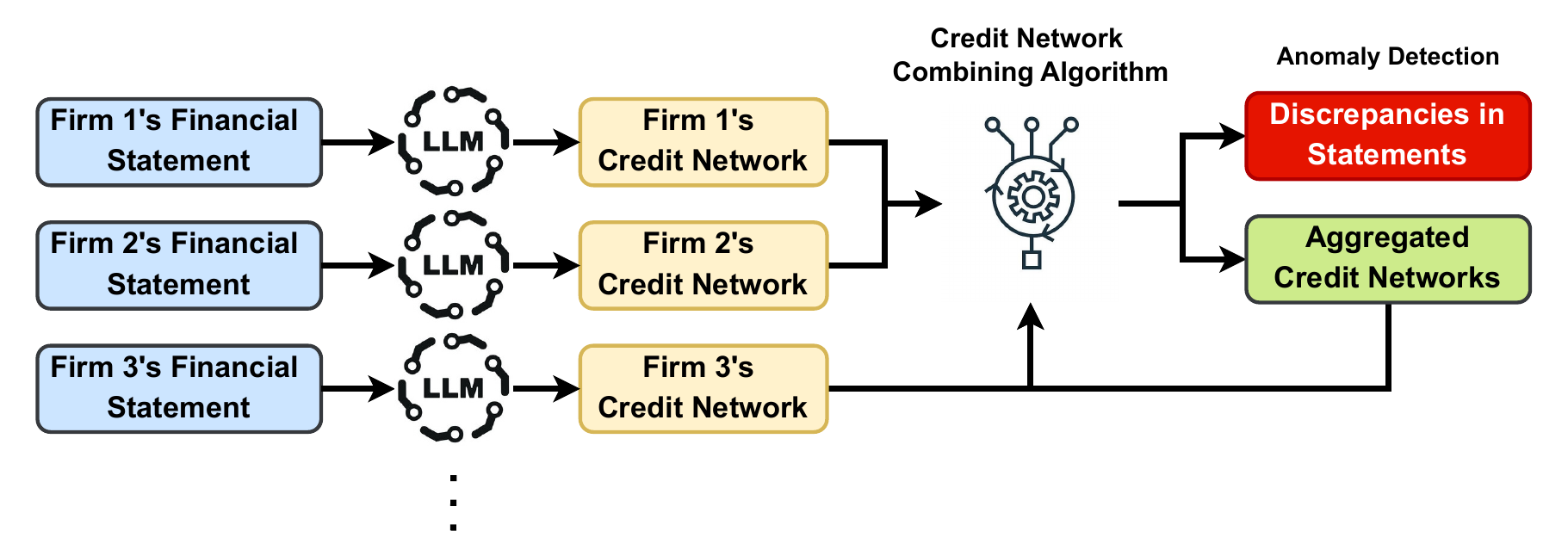}
  \caption{Workflow for translating financial statements to credit networks.}
  \label{fig:stage1_framework}
\end{figure*}

\section{Format of Financial statements}

A financial statement is an official record that outlines the financial activities and position of a business, individual, or other entity, typically over a specific period. 
For our purposes, we focus on a simplified version of the financial statement, which includes key elements that represent the financial position such as assets, liabilities, and shareholders' equity (e.g., retained earnings). 
These components are usually presented in a clear balance sheet format, following GAAP or IFRS standards, with additional explanatory text. 
As such, a financial statement can be seen as a natural language description of a balance sheet.
An example of the format of financial statement we use is shown below.

\begin{framed}
\textbf{Firm A – Annual Financial Disclosure (2024)}

As of December 31, 2024, Firm A maintains \$15 million in liquid assets, primarily held across cash deposits at Bank X (\$10 million) and short-term Treasury securities (\$5 million). 
However, \$2 million of the cash reserves are currently encumbered as collateral for a standby letter 
of credit and are unavailable for general use.

In terms of liabilities, Firm A has entered into a revolving credit facility with Firm B, carrying an 
outstanding balance of \$5 million. Additionally, a term loan agreement with Firm C requires repayment 
of \$4 million by mid-2025. Trade financing arrangements with Firm D have resulted in \$2 million of 
accounts payable.

On the receivables side, Firm E has executed a promissory note to Firm A for \$3 million, 
though payment timelines remain subject to performance milestones. 
Firm F acknowledges an outstanding \$1 million trade receivable, but collection efforts are 
ongoing due to disputed service terms.
\end{framed}

We gather financial statements from two sources: real-world financial reports and those generated by LLMs. 
The real-world reports are sourced from the official websites of publicly listed companies, and we extract the components that present key financial information to construct our financial statements.
Besides, to evaluate our framework across credit networks with different topologies and numbers of firms, we also use LLMs to generate financial statements. 
These synthetic statements are crafted to closely replicate the format and structure of real-world reports, while reflecting a range of network topologies and scales.

\section{Translation to Credit Networks}

\subsection{Framework}
We employ LLMs to translate the financial statement of each firm into a credit network that pertains solely to that firm, and then aggregate the individual credit networks to form a set of system-wide credit networks\footnote{This results in a set rather than a single unified network because some firms may not have debt relationships with others in the system.}.
During the integration process, inconsistencies across firms’ statements will be automatically detected.
For example, if Firm A reports owing \$5 million to Firm B, but Firm B claims that A owes \$6 million, a discrepancy is flagged and human intervention is required. 
There are various ways in which human involvement can occur when discrepancies arise. For example, an expert could investigate the issue, resolve it, and then resume the translation process. 
In this study, however, we do not delve into these scenarios. 
Instead, our method will simply trigger an alert to indicate that a discrepancy has occurred.
The workflow of our framework is illustrated in Figure~\ref{fig:stage1_framework} and
the complete procedure is outlined in Algorithm~\ref{alg:credit-network-integration}.

\begin{algorithm}[H]
\caption{Iterative Integration of Credit Networks}
\label{alg:credit-network-integration}
\begin{algorithmic}[1]
\STATE Initialize aggregated credit networks $\mathcal{C}_{\text{agg}} \gets \emptyset$
\FOR{each financial statement $S_i$}
    \STATE Translate $S_i$ into credit network $\mathcal{C}_i$ using the LLM
    \STATE Perform anomaly detection between $\mathcal{C}_i$ and $\mathcal{C}_{\text{agg}}$
    \IF{anomaly detected}
        \STATE Trigger alert for human intervention
        \STATE \textbf{break} \COMMENT{Exit the loop due to detected discrepancy}
    \ENDIF
    \STATE Integrate $\mathcal{C}_i$ into $\mathcal{C}_{\text{agg}}$
\ENDFOR
\STATE \textbf{return} Final aggregated credit networks $\mathcal{C}_{\text{agg}}$
\end{algorithmic}
\end{algorithm}

\subsection{Prompts for LLMs}

The prompts in our framework contains two major components.
The first component introduces the representation of credit networks to the LLM. 
The second component instructs the LLM to extract the firm's name, external assets, and liabilities from the financial statement and summarize them in a predefined format that supports further aggregation.
These two components are concatenated as the prompt inputs to the LLM.

\begin{framed}
\textbf{Prompts for introducing credit networks:} 

You are given a credit network represented in adjacency matrix L, where L[i][j] represents the amount that firm i owes firm j.
An external asset vector e, where e[i] is the amount of external (non-network) assets held by firm i.

Here is an example:

L = [[0, 5, 0, 0],
[0, 0, 4, 0],
[4, 0, 0, 5],
[9, 0, 0, 0]]

e = [3, 4, 5, 6]


\end{framed}

\begin{framed}
\textbf{Prompts for extracting information:} 

You are a financial data extraction assistant for constructing a credit network model.

Given a financial statement, extract:
\begin{enumerate}
    \item The reporting firm's name;
    \item The reporting firm's external assets (in millions). External assets are assets held by the reporting firm that originate from entities outside the network, such as deposits at commercial banks, government bonds, real estate, etc;
    \item A list of all liabilities, specifying:
        \begin{itemize}
            \item borrower name;
            \item lender name;
            \item amount (in millions).
        \end{itemize}
\end{enumerate}

The reporting firm can be either a borrower or a lender. List all relationships explicitly.

Output Python code in the following format:
\begin{enumerate}
    \item firm = "<reporting firm name>"
    \item external\_assets = <amount>
    \item liabilities = [(<borrower>, <lender>, <amount>), $\dotsc$]
\end{enumerate}

Here is the financial statement:

<INSERT\_FINANCIAL\_STATEMENT\_HERE>

\end{framed}

\section{Identifying Execution Strategies}

\begin{figure}[htbp]
  \centering
  \includegraphics[width=0.5\textwidth]{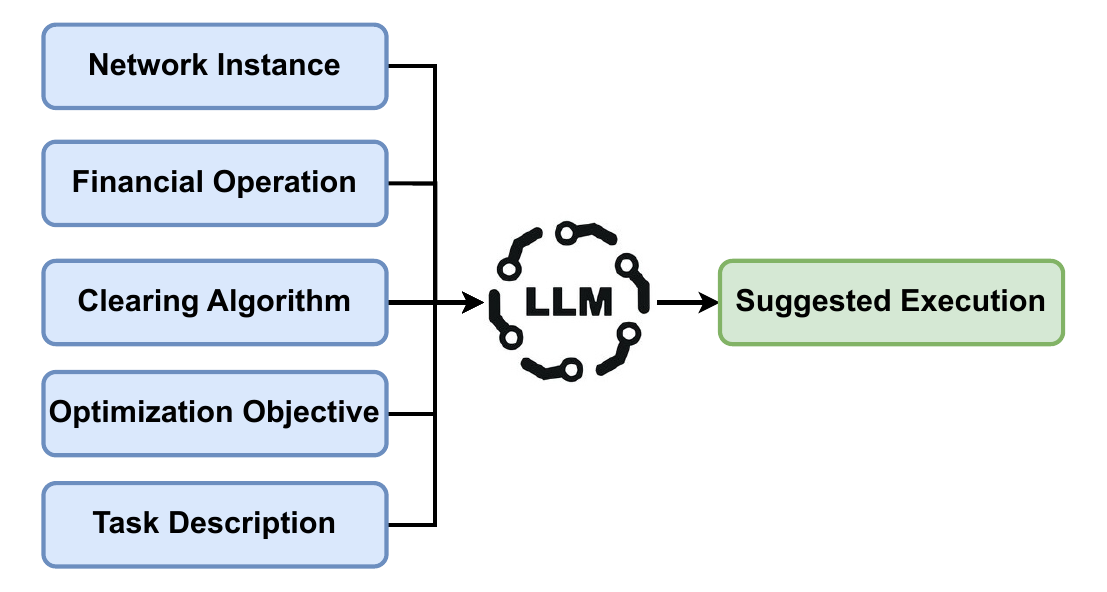}
  \caption{Inputs and outputs of LLMs for identifying effective execution strategies.}
  \label{fig:stage2_framework}
\end{figure}

We explore the capabilities of LLMs in analyzing credit networks and determining effective execution strategies for financial operations. 
As illustrated in Figure~\ref{fig:stage2_framework}, we prompt an LLM with the following components:
\begin{itemize}
\item \textbf{Credit network instance:} A matrix representation of the credit network, accompanied by an explanation of its structure and interpretation;
\item \textbf{Financial operation description:} A description of the financial operation and its impact on the credit network;
\item \textbf{Clearing algorithm:} A Python implementation of a clearing algorithm applicable to the matrix representation of the credit network;
\item \textbf{Optimization objective:} A target criterion for the financial operation, such as maximizing total assets;
\item \textbf{Task Description:} A description of the task to find an effective execution strategy that maximizes the objective after clearance.
\end{itemize}
The LLM then generates a proposed execution strategy for the financial operation, along with an explanation of its reasoning process.

\section{Experimental Results}

\subsection{Translation to Credit Networks}\label{sec: translation results}

To demonstrate that our framework can translate financial statements reflecting various credit network structures, we begin by generating a diverse set of credit networks.
Using the liabilities and external assets within these networks, we then construct synthetic financial statements. 
The credit networks are modeled to follow five distinct topologies:
\begin{enumerate}
    \item \textbf{Erd\H{o}s--R\'enyi:} This represents a random network where the edges, liabilities, and external assets are randomly generated;
    \item \textbf{Core-Periphery:} In a Core-Periphery topology, a small group of core institutions (e.g., large banks) are densely interconnected and handle the majority of credit flows, while peripheral nodes, such as smaller banks or firms, are sparsely connected and primarily interact with the core rather than with each other. This configuration is common in interbank markets and reflects a hub-and-spoke pattern, with the core acting as central hubs for liquidity and financial services;
    \item \textbf{Isolated Blocks:} This topology consists of distinct clusters of nodes that engage in credit relationships primarily within their own group, with no connections to other clusters;
    \item \textbf{DAG-SSCs:} A DAG-SCCs network comprises multiple strongly connected components (SCCs), with firms within each SCC being fully interconnected, while connections between firms in different SCCs are sparse;
    \item \textbf{German Banks:} This network topology is derived from the financial statements of twenty-two German banks~\cite{chen2021financial}. The network consists of twenty-three nodes: twenty-two nodes represent the individual banks, and an additional node represents all other banks in the banking system.
\end{enumerate}
For each of these topologies (except the German Banks), we create ten credit network instances (possibly with isolated blocks) and each of them has ten firms with randomly generated debt relationships, liabilities, and external assets.
The parameters for generating these credit networks are shown in Table~\ref{tab: params}.
For each of these instances, we then use the general-purpose LLM GPT-4o~\cite{openai2024gpt4o} to generate the corresponding financial statements for the involved firms using the liabilities and external assets in the credit network.
The translation of these financial statements into a set of credit networks forms a \newterm{translation task}.
Additionally, we apply the same approach to generate financial statements, intentionally introducing inconsistencies to serve as test cases for the anomaly detection module.

\begin{table}[!h]
    \centering
    \begin{tabular}{lc}
    \toprule
     \textbf{Parameters}    &  \textbf{Values} \\
     \midrule
     \textbf{Liabilities}    &  Uniform[15, 40]\\
     \textbf{External Assets}    &  Uniform[30, 50]\\
     \bottomrule
    \end{tabular}
    \caption{Network generation parameters.}
    \label{tab: params}
\end{table}

Table~\ref{Table: Translation Results} reports the performance of our framework with GPT-4o and a single prompting on solving translation tasks across various credit network topologies. 
For each topology, we report the number of successful translations out of the total number of tasks, both with and without intentionally introduced inconsistencies. 
We find that regardless of network topologies, our framework can accurately translate financial statements in the absence of inconsistencies and reliably detect all introduced inconsistencies, triggering alerts for human review.

We further evaluate the scalability of our framework by substantially increasing the number of firms in the financial system. 
We find that the framework can reliably handle systems with over 2,000 firms. 
Importantly, the primary limitation at this scale stems from the context window size of the LLM, not from our framework itself. 
This highlights the strong scalability of our approach.

\begin{table}[!htbp]
  \centering
  \begin{tabular}{lcc}
    \toprule
    \textbf{Topologies} & \textbf{w.o Inconsistency} & \textbf{w. Inconsistency}\\
    \midrule
    \textbf{Erd\H{o}s--R\'enyi} & 10 / 10 & 10 / 10\\
    \midrule
    \textbf{Core-Periphery} & 10 / 10 & 10 / 10\\
    \midrule
    \textbf{Isolated Blocks} & 10 / 10 & 10 / 10\\
    \midrule
    \textbf{DAG-SCCs} & 10 / 10 & 10 / 10\\
    \midrule
    \textbf{German Banks}& 1 / 1 & NA\\
    \bottomrule
  \end{tabular}
  \caption{Results of financial statement translation.}
  \label{Table: Translation Results}
\end{table}

In addition to using synthetic data, we also test our framework with real-world financial statements of publicly listed firms.
For this purpose, we select four companies: AT\&T, DirecTV, Apple, and Broadcom, which follow the isolated block topology, where AT\&T has a debt relationship with DirecTV and Apple has one with Broadcom. 
Rather than using the entire financial statements, we extract key components that provide essential financial information to construct the credit network\footnote{The financial statements and credit networks used in these experiments will be shared upon acceptance of the paper, in accordance with the anonymous review policy.}. 
We then manually assess the framework's output, and our findings show that it successfully translates these financial statements.
Although the resulting credit network is small, the experiment demonstrates that our approach is applicable to real-world data. 
Combined with its demonstrated scalability, this suggests that our method can be extended to larger real-world financial systems.

\begin{table*}[!htbp]
  \centering
  \begin{tabular}{llcccccccccc}
    \toprule
    \textbf{Topologies} & \textbf{Execution Strategies} & \multicolumn{10}{c}{\textbf{Network Instances}} \\
    \midrule
    &  & \textbf{ER1} & \textbf{ER2} & \textbf{ER3} & \textbf{ER4} & \textbf{ER5} & \textbf{ER6} & \textbf{ER7} & \textbf{ER8} & \textbf{ER9} & \textbf{ER10} \\
    \multirow{4}{*}{\textbf{Erd\H{o}s--R\'enyi}}  & No Compression     & 79.99  & 120.01 & 59.99  & 90.02  & 96.84  & 88.01  & 58.02  & 80.02  & 94.02  & 64.01 \\
                                                  & Random Compression & 122.91 & 153.35 & 106.75 & 129.26 & 148.28 & 149.45 & 102.16 & 108.12 & 124.95 & 98.32 \\
                                                  & Heuristic Baseline & 158.81 & 249.38 & 137.00 & 169.43 & 168.25 & 176.00 & 124.00 & 131.00 & 179.00 & 118.00 \\
                                                  & LLM Suggestion     & \textbf{198.01} & \textbf{319.01} & \textbf{236.25} & \textbf{258.50} & \textbf{285.01} & \textbf{277.01} & \textbf{251.01} & \textbf{246.38} & \textbf{327.89} & \textbf{233.01} \\
    \midrule
    &  & \textbf{CP1} & \textbf{CP2} & \textbf{CP3} & \textbf{CP4} & \textbf{CP5} & \textbf{CP6} & \textbf{CP7} & \textbf{CP8} & \textbf{CP9} & \textbf{CP10} \\
    \multirow{4}{*}{\textbf{Core-Periphery}} & No Compression     & 55.01  & 88.69  & 103.61 & 105.98 & 71.63  & 176.27 & 72.99  & 35.99  & 115.18 & 83.99 \\
                                             & Random Compression & 98.08  & 112.96 & 124.34 & 125.05 & 86.25  & 178.09 & 120.68 & 82.15  & 120.95 & 125.36 \\
                                             & Heuristic Baseline & 125.66 & 134.03 & 151.31 & 153.54 & 96.40  & 180.75 & 135.93 & 94.62  & 158.72 & 146.40 \\
                                             & LLM Suggestion     & \textbf{149.99} & \textbf{153.29} & \textbf{198.59} & \textbf{185.04} & \textbf{108.83} & \textbf{190.57} & \textbf{149.50} & \textbf{141.37} & \textbf{189.99} & \textbf{203.86} \\
    \midrule
    &  & \textbf{IB1} & \textbf{IB2} & \textbf{IB3} & \textbf{IB4} & \textbf{IB5} & \textbf{IB6} & \textbf{IB7} & \textbf{IB8} & \textbf{IB9} & \textbf{IB10} \\
    \multirow{4}{*}{\textbf{Isolated Blocks}} & No Compression     & 145.01 & 55.99  & 45.99  & 115.01 & 86.01  & 21.99  & 48.01  & 62.01  & 66.01  & 101.02 \\
                                              & Random Compression & 158.45 & 85.03  & 90.78  & 126.85 & 95.54  & 145.23 & 91.49  & 116.94 & 84.29  & 114.98 \\
                                              & Heuristic Baseline & 170.00 & 128.00 & 129.00 & 149.00 & 121.00 & 175.00 & 125.00 & 140.88 & 106.78 & 137.00 \\
                                              & LLM Suggestion     & \textbf{212.01} & \textbf{223.01} & \textbf{200.50} & \textbf{202.50} & \textbf{157.70} & \textbf{205.50} & \textbf{187.01} & \textbf{191.38} & \textbf{161.12} & \textbf{165.01} \\
    \midrule
    &  & \textbf{SCC1} & \textbf{SCC2} & \textbf{SCC3} & \textbf{SCC4} & \textbf{SCC5} & \textbf{SCC6} & \textbf{SCC7} & \textbf{SCC8} & \textbf{SCC9} & \textbf{SCC10} \\
    \multirow{4}{*}{\textbf{DAG-SCCs}}    & No Compression     & 188.49 & 161.53 & 74.73  & 84.29  & 100.13 & 95.79  & 95.86  & 98.74  & 107.44 & 89.07 \\
                                          & Random Compression & 188.58 & 166.94 & 117.37 & 98.85  & 113.86 & 102.56 & 114.34 & 112.01 & 118.02 & 99.03 \\
                                          & Heuristic Baseline & 192.05 & 170.73 & 131.22 & 110.82 & 130.79 & 113.78 & 131.57 & 120.48 & 125.78 & 104.32  \\
                                          & LLM Suggestion     & \textbf{197.66} & \textbf{184.63} & \textbf{134.72} & \textbf{125.81} & \textbf{133.03} & \textbf{131.31} & \textbf{138.01} & \textbf{131.33} & \textbf{138.09} & \textbf{109.68} \\
    \bottomrule
  \end{tabular}
  \caption{The sum of total assets resulted from different portfolio compression strategies using GPT-4o and a single prompting.}
  \label{Table1: Portfolio Compression Results}
\end{table*}

\begin{table*}[htbp]
  \centering
  \begin{tabular}{ll*{10}{c}}
    \toprule
    \textbf{Topologies} & \textbf{Execution Strategies} & \multicolumn{10}{c}{\textbf{Network Instances}} \\
    \midrule
    &  & \textbf{ER1} & \textbf{ER2} & \textbf{ER3} & \textbf{ER4} & \textbf{ER5} & \textbf{ER6} & \textbf{ER7} & \textbf{ER8} & \textbf{ER9} & \textbf{ER10} \\
    \multirow{4}{*}{\textbf{Erd\H{o}s--R\'enyi}}  & No Removal         & 110.01 & 110.01 & 55.86 & 78.01  & 58.01  & 64.02  & 87.99  & 52.01  & 114.03 & 87.79 \\
                                                  & Random Removal     & 99.21  & 136.43 & 44.16 & 66.86  & 45.09  & 62.58  & 73.95  & 40.44  & 102.55 & 80.40 \\
                                                  & Heuristic Baseline & 111.01 & 144.41 & 59.76 & 103.82 & 81.06  & 65.11  & 90.64  & 185.51 & 164.34 & 188.50 \\
                                                  & LLM Suggestion     & \textbf{334.01} & \textbf{359.01} & \textbf{96.18} & \textbf{215.01} & \textbf{231.01} & \textbf{165.44} & \textbf{197.01} & \textbf{196.01} & \textbf{248.50} & \textbf{210.01} \\
    \midrule
    &  & \textbf{CP1} & \textbf{CP2} & \textbf{CP3} & \textbf{CP4} & \textbf{CP5} & \textbf{CP6} & \textbf{CP7} & \textbf{CP8} & \textbf{CP9} & \textbf{CP10} \\
    \multirow{4}{*}{\textbf{Core-Periphery}} & No Removal         & 79.99  & 58.01  & 155.39 & 119.52 & 182.54  & 108.14 & 66.02  & 76.02  & 59.99  & 182.08 \\
                                             & Random Removal     & 64.62  & 51.92  & 111.34 & 91.21  & 133.82  & 100.44 & 61.24  & 69.11  & 53.95  & 45.52 \\
                                             & Heuristic Baseline & 85.43  & 97.63  & 166.14 & 153.45 & 197.18  & 114.28 & 244.01 & 136.22 & 68.26  & 190.17 \\
                                             & LLM Suggestion     & \textbf{222.01} & \textbf{211.02} & \textbf{246.01} & \textbf{178.01} & \textbf{275.01}  & \textbf{172.02} & \textbf{244.01} & \textbf{211.01} & \textbf{117.90} & \textbf{223.01} \\
    \midrule
    &  & \textbf{IB1} & \textbf{IB2} & \textbf{IB3} & \textbf{IB4} & \textbf{IB5} & \textbf{IB6} & \textbf{IB7} & \textbf{IB8} & \textbf{IB9} & \textbf{IB10} \\
    \multirow{4}{*}{\textbf{Isolated Blocks}} & No Removal         & 41.99  & 67.99  & 35.99  & 158.02 & 45.99  & 107.02 & 20.01 & 62.01  & 135.02 & 58.0 \\
                                              & Random Removal     & 40.81  & 70.77  & 30.78  & 64.40  & 48.83  & 31.66  & 21.26 & 55.68  & 29.13  & 46.79 \\
                                              & Heuristic Baseline & 132.99 & 169.01 & 139.99 & 259.01 & 161.99 & 189.01 & 57.50 & 141.01 & 253.01 & 152.01 \\
                                              & LLM Suggestion     & \textbf{172.01} & \textbf{222.01} & \textbf{226.01} & \textbf{259.01} & \textbf{166.02} & \textbf{210.01} & \textbf{63.01} & \textbf{217.01} & \textbf{253.01} & \textbf{241.01} \\
    \midrule
    &  & \textbf{SCC1} & \textbf{SCC2} & \textbf{SCC3} & \textbf{SCC4} & \textbf{SCC5} & \textbf{SCC6} & \textbf{SCC7} & \textbf{SCC8} & \textbf{SCC9} & \textbf{SCC10} \\
    \multirow{4}{*}{\textbf{DAG-SCCs}} & No Removal            & 138.52 & 134.08 & 134.93 & 87.36  & 87.23  & 68.15  & 106.03 & 95.49  & 159.01 & 103.03 \\
                                          & Random Removal     & 76.51  & 87.29  & 78.69  & 66.77  & 57.05  & 55.58  & 77.43  & 58.64  & 73.65  & 73.81 \\
                                          & Heuristic Baseline & 139.31 & 143.10 & 162.05 & 123.78 & 101.22 & 71.12  & 148.50 & 98.28  & 161.93 & 106.26 \\
                                          & LLM Suggestion     & \textbf{152.23} & \textbf{148.50} & \textbf{165.75} & \textbf{135.88} & \textbf{124.51} & \textbf{116.51} & \textbf{167.51} & \textbf{156.01} & \textbf{176.25} & \textbf{146.51} \\
    \bottomrule
  \end{tabular}
  \caption{The sum of total assets resulted from different debt removal strategies using GPT-4o and a single prompting.}
  \label{Table2: Debt Removal Results}
\end{table*}


\subsection{LLM-Suggested Execution}

\subsubsection{Portfolio Compression.}

For portfolio compression, we consider the following execution strategies as baseline comparisons:

\begin{enumerate}
\item \textbf{No Compression:} No debt cycles are compressed;
\item \textbf{Random Compression:} A randomly selected subset of debt cycles is compressed;
\item \textbf{Greedy Strategy:} Compresses the debt cycles with the highest \newterm{weighted flow}, defined as the product of the cycle’s minimum liability and the number of firms involved;
\end{enumerate}

The \textbf{No Compression} baseline reflects the default scenario where no operation occurs, serving as a reference point to ensure that any proposed strategy yields improvement over inaction. 
The \textbf{Random Compression} baseline illustrates the performance of unstructured operations. 
Lastly, the \textbf{Greedy Strategy} targets the most impactful cycles, compressing the top three that maximize the weighted flow, thereby aiming to deliver the greatest systemic benefit.

In Table~\ref{Table1: Portfolio Compression Results}, we present the sum of firms' total assets (i.e., $\sum_{i\in [n]}a_i$) after applying the LLM-suggested executions, compared to those produced by the baseline strategies. 
To ensure diversity in network structures, we use the topologies described in Section~\ref{sec: translation results}. 
For each topology, we generate ten network instances, each labeled with the name of its corresponding topology.

Across all topologies in Table~\ref{Table1: Portfolio Compression Results}, the No Compression strategy consistently yields the poorest results, while Random Compression provides a modest improvement. 
This suggests that, although it is possible to construct credit networks where portfolio compression is detrimental, it generally enhances system performance. 
Note that the effectiveness of random strategy depends heavily on the nature of the financial operations involved. 
As we show later for debt removal, random strategy may harm the system.

Furthermore, both the heuristic and LLM-based strategies outperform the No Compression and Random Compression baselines, with our LLM approach delivering the strongest overall performance. Interestingly, we find that compressing a larger number of debt cycles does not always result in better outcomes. 
Instead, the LLM approach appears to selectively target specific debt cycles, effectively enhancing overall network performance.

\subsubsection{Debt Removal}

To evaluate the impact of debt removal, we compare against the following baseline strategies:
\begin{enumerate}
\item \textbf{No Removal:} Debts are left unchanged; 
\item \textbf{Random Removal:} A randomly selected subset of debts is removed;
\item \textbf{Greedy Strategy:} Debts are selectively removed for a subset of insolvent firms such that:
(1) the target firms become solvent,
(2) the total amount of debt removed is minimized, and
(3) the removed debt does not exceed the recovered external assets net of default costs.
\end{enumerate}

From Table~\ref{Table2: Debt Removal Results}, it is evident that, unlike the portfolio compression results, the Random strategy produces the worst outcomes, rather than the No Operation strategy. 
This occurs because indiscriminately removing debt negatively impacts the total assets, with the detrimental effect increasing as more debt is removed. 
Similar to the portfolio compression results, both the heuristic and LLM-based strategies outperform both No Removal and Random Removal, with the LLM approach achieving the best overall performance. 
Meanwhile, the LLM strategy also provides insight into its reasoning process. 
For example, the LLM strategy recognizes that removing large amounts of debt can damage the total assets of the credit network. As a result, it tends to identify insolvent firms and remove only a small amount of debt to restore solvency, demonstrating that the reasoning behind the LLM strategy is sound.

Finally, beyond the results presented in Table~\ref{Table1: Portfolio Compression Results} and Table~\ref{Table2: Debt Removal Results}, we conduct the same experiments on German banks for both financial operations. 
The outcomes align with those obtained using synthetic data, demonstrating the effectiveness of our approach in real-world settings.

\section{Conclusion and Discussion}

We explore the applications of LLMs to financial systems represented by credit networks.
Our contributions are twofold. 
First, we present a framework that translates unstructured financial statements into their corresponding credit network representations. We show that this framework is scalable and adaptable to financial systems with diverse underlying structures. 
Second, we demonstrate that LLMs can reason effectively about the execution of financial operations within these networks. 
These findings lay the groundwork for developing specialized LLMs tailored to financial system analysis.

For future research, we aim to explore the capabilities of LLMs in recommending the execution of financial operations without limiting the scope to specific operation types. 
In practice, multiple financial operations may need to be executed in combination to enhance the overall performance of a financial system, introducing a complex combinatorial optimization challenge. 
Thus, even if LLMs can identify suboptimal but effective execution strategies in such scenarios, it would represent a significant contribution.

\bibliographystyle{ACM}
\bibliography{ref}

\end{document}